\documentclass[draft]{agujournal2019}
\usepackage{url}
\usepackage{amsmath,amssymb}
\usepackage{lineno}
\usepackage[inline]{trackchanges}
\usepackage{soul}

\newcommand {\Dt}[1]{\frac{D #1}{D t}}
\newcommand {\bv}{\boldsymbol{v}}
\newcommand {\bOmega}{\boldsymbol{\Omega}}

\newcommand {\eR}{\mathbf{e_r}}
\newcommand {\ey}{\mathbf{e_y}}
\newcommand {\ex}{\mathbf{e_x}}

\newcommand {\pR}[1]{\frac{\partial #1}{\partial r}}

\draftfalse

\journalname{Geophysical Research Letters}

\begin{document}

\title{Slantwise convection and heat transport in icy moon oceans}

\authors{Yaoxuan Zeng\affil{1} and Malte F. Jansen\affil{1}}

\affiliation{1}{Department of the Geophysical Sciences, The University of Chicago, Chicago, IL 60637, USA}

\correspondingauthor{Yaoxuan Zeng}{yxzeng@uchicago.edu}

\begin{keypoints}
\item Slantwise convection aligned with the rotation axis is expected to occur in icy moon oceans
\item Scaling laws for heat transport by slantwise convection are developed
\item Slantwise convection drives poleward heat transport and may support the poleward-thinning ice shell observed on Enceladus
\end{keypoints}

\begin{abstract}

Ocean heat transport on icy moons shapes the ice shell topography, a primary observable of these moons. Two key processes control the heat transport: baroclinic instability driven by surface buoyancy contrasts and convective instability driven by heating from the core. However, global ocean simulations cannot accurately resolve convection under realistic icy moon conditions and instead often use Earth-based convective parameterizations, which capture only vertical convective mixing and cannot represent rotation-aligned slantwise convection on icy moons. We use high-resolution convection-resolving simulations to investigate ocean heat transport by slantwise convection in a parameter regime relevant to icy moons, isolated from baroclinic instability. Total heat transport follows the Coriolis–Inertial–Archimedean scaling with an added latitude dependence. The vertical transport increases with latitude, and the meridional transport is poleward. These results indicate that slantwise convection redistributes heat toward the poles, favoring a poleward-thinning ice shell, qualitatively consistent with Enceladus’s observed ice thickness distribution.

\end{abstract}

\section*{Plain Language Summary}

Ocean heat transport on icy moons like Europa and Enceladus plays a key role in shaping the thickness distribution of the overlying ice shell, one of the primary observable features of these moons. Two main processes are thought to control heat transport in their oceans: large-scale eddies driven by baroclinic instability, and small-scale convection driven by heating from the seafloor. However, simulating both processes together in global ocean models is computationally too expensive under icy moon conditions. Existing parameterizations for convective heat transport are based on Earth’s oceans, where convection and the associated heat flux are assumed to be purely vertical. On icy moons, however, slantwise convection tends to align with the planet’s rotation axis, making Earth-based parameterizations inaccurate. In this study, we use localized high-resolution numerical simulations to investigate slantwise convection and its role in ocean heat transport under icy moon conditions. Our results show that slantwise convection redistributes heat toward the poles, favoring a thinner ice shell at higher latitudes, qualitatively consistent with the observed ice thickness distribution on Enceladus.

\section{Introduction}

Strong evidence supports the existence of global subsurface oceans on several icy moons \cite{khurana1998induced,neubauer1998sub,zimmer2000subsurface,kivelson2002permanent,iess2012tides,mckinnon2015effect,thomas2016enceladus}. These oceans, in contact with rocky cores, may support ongoing water–rock interactions and methanogenesis, making icy moons prime targets in the search for extraterrestrial life in the solar system \cite{mckay2008possible,hsu2015ongoing,sekine2015high,waite2017cassini}. Ocean circulation plays an essential role in setting the habitability and the detectability of biosignatures on these moons. However, because direct observations of ocean dynamics are lacking, researchers rely on surface measurements, particularly ice shell topography, to infer subsurface ocean circulation and heat transport. If the ice shell is in a quasi-equilibrium state, then basal melting and freezing must balance the mass transport of ice flow. The observed ice thickness gradients, which drive ice flow \cite{vcadek2019long,shibley2024infer}, are therefore intimately connected to ocean circulation and heat transport through basal melting and freezing \cite{kang2022does,zeng2024effect}.

Global General Circulation Models (GCMs) of icy moon oceans need to simulate both convection driven by heating from the rocky core and baroclinic eddies driven by meridional buoyancy gradients at the ice–ocean interface \cite{zeng2021ocean,ashkenazy2021dynamic,kang2022does,kang2023modulation,zeng2024effect,ames2025ocean}. Convection in icy moon oceans is constrained by rotation and organizes into columns aligned with the planetary rotation axis, known as slantwise convection \cite{ashkenazy2021dynamic,zeng2021ocean,bire2022exploring,kang2023modulation,zeng2025symmetric}. Surface buoyancy gradients can arise from salinity changes due to freezing and melting, or from pressure-dependent freezing point variations linked to ice shell topography \cite{fofonoff1983algorithms,jackett1995minimal}. These gradients drive baroclinic instability, generating eddies that transport heat from warm, thin-ice regions to cold, thick-ice regions \cite{kang2022icy,kang2022different,zeng2024effect,zhang2024ocean}.

Convective plumes under icy moon conditions are extremely small, about 0.01--0.1$^\circ$ in latitude (Table~S1), and thus not resolved in current GCMs. Most models, therefore, rely on convective adjustment schemes developed for Earth's oceans \cite{marotzke1991influence}, which assume purely vertical mixing (i.e., parallel to gravity). This assumption is incompatible with the rotationally aligned convection on icy moons \cite{zeng2025symmetric}, raising concerns about the accuracy of current parameterizations. It is therefore essential to characterize heat transport by slantwise convection and develop more appropriate schemes for global models.

The simplest abstraction for ocean convection is Rayleigh–Bénard convection, which models buoyancy-driven flow between two parallel plates heated from below and cooled from above. Rotational effects are often studied using an $f$-plane, where the rotation axis is aligned with gravity and perpendicular to the plates (see \citeA{ecke2023turbulent} and references therein). In the rapidly rotating, turbulent regime, assuming a dominant balance among the Coriolis force, inertial acceleration, and buoyancy force leads to a CIA (Coriolis–Inertial–Archimedean) scaling (c.f. \citeA{aurnou2020connections} and references therein) for the flux-gradient relation:

\begin{equation}\label{eq:diffusive-free-heat}
    |\nabla b| = \Delta b/H \propto B^{2/5} (2\Omega)^{4/5} H^{-4/5},
\end{equation}

\noindent where $\nabla b$ is the buoyancy gradient, $\Delta b = g \alpha_T \Delta T$ is the buoyancy contrast between the two plates, $B = (Q g\alpha_T)/(\rho_0 c_p)$ is the buoyancy flux, $\Omega$ is the rotation rate, $H$ is the depth scale, and $g$, $\alpha_T$, $\Delta T$, $Q$, $\rho_0$, and $c_p$ are gravity, thermal expansion coefficient, temperature contrast, heat flux, reference density, and heat capacity, respectively. This scaling is consistent with the scaling law in the diffusion-free limit, also known as the ultimate regime, for rapidly rotating fluids, where planetary rotation is sufficiently strong and the flow is sufficiently turbulent that the effects of molecular viscosity and diffusivity become negligible (c.f. \citeA{julien2012heat, gastine2016scaling}).

On planetary scales, the rotation axis is often tilted relative to gravity. To capture this, many studies adopt a tilted $f$-plane, where the rotation vector has both vertical and horizontal components \cite{flasar1978turbulent,hathaway1979convective,hathaway1980convective,hathaway1983three,von2015generation,novi2019rapidly,currie2020convection,tro2024parameterized}. The horizontal component of the rotation vector breaks horizontal symmetry and stretches convective structures along the rotation axis. At high rotation rates and large thermal contrasts, strong zonal flows can emerge, shearing the plumes and causing intermittent convection near the equator \cite{von2015generation}.

Recent studies have also explored Rayleigh–Bénard convection in spherical shells \cite{soderlund2014ocean,gastine2016scaling,soderlund2019ocean,amit2020cooling,wang_diffusionfree_2021,kvorka2022numerical,bire2022exploring,lemasquerier2023europa,gastine2023latitudinal,hartmann2024toward,fan2024scaling}. Several of these studies have examined whether bottom heat is preferentially transported toward the poles or the equator \cite{soderlund2019ocean,amit2020cooling,kvorka2022numerical,bire2022exploring,lemasquerier2023europa}. The parameters of these simulations remain far from those relevant to icy moons, so their results should be interpreted with caution when applied to real ocean conditions. For example, if the viscosity ($\nu$) is set too high for numerical stability, leading to a high Ekman number ($\mathrm{Ek} = \nu\Omega^{-1} H^{-2}$), high-latitude convective plumes can be artificially suppressed. This suppression may not occur in the low-viscosity (small $\mathrm{Ek}$) regime that likely best represents icy moon ocean interiors. In this low-viscosity limit, \citeA{bire2022exploring} found enhancement of vertical heat transport toward the poles under uniform bottom heating.

Slantwise convection drives heat transport both vertically and meridionally. Theoretical studies have shown that, in rapidly rotating regimes, the heat transport associated with the most unstable linear mode is aligned with the rotation axis, implying transport in both directions \cite{flasar1978turbulent,hathaway1979convective}. However, most studies of rotating Rayleigh–Bénard convection have focused on vertical transport, with the meridional component less explored. In this work, we conduct convection-resolving simulations to study the relationship between buoyancy flux and buoyancy gradient in icy moon oceans, considering heat transport in both directions. We target a parameter regime comparable to real planetary conditions on icy moons. Section~\ref{sec:numerical-method} describes our numerical method. Section~\ref{sec:scaling} presents simulation results and scaling laws. Section~\ref{sec:implication} discusses the implications for ocean heat transport and ice shell topography on icy moons. Section~\ref{sec:discussion} provides discussion and concluding remarks.

\section{Numerical method}\label{sec:numerical-method}

\begin{figure}[t!]
    \centering
    \includegraphics[width=1.0\linewidth]{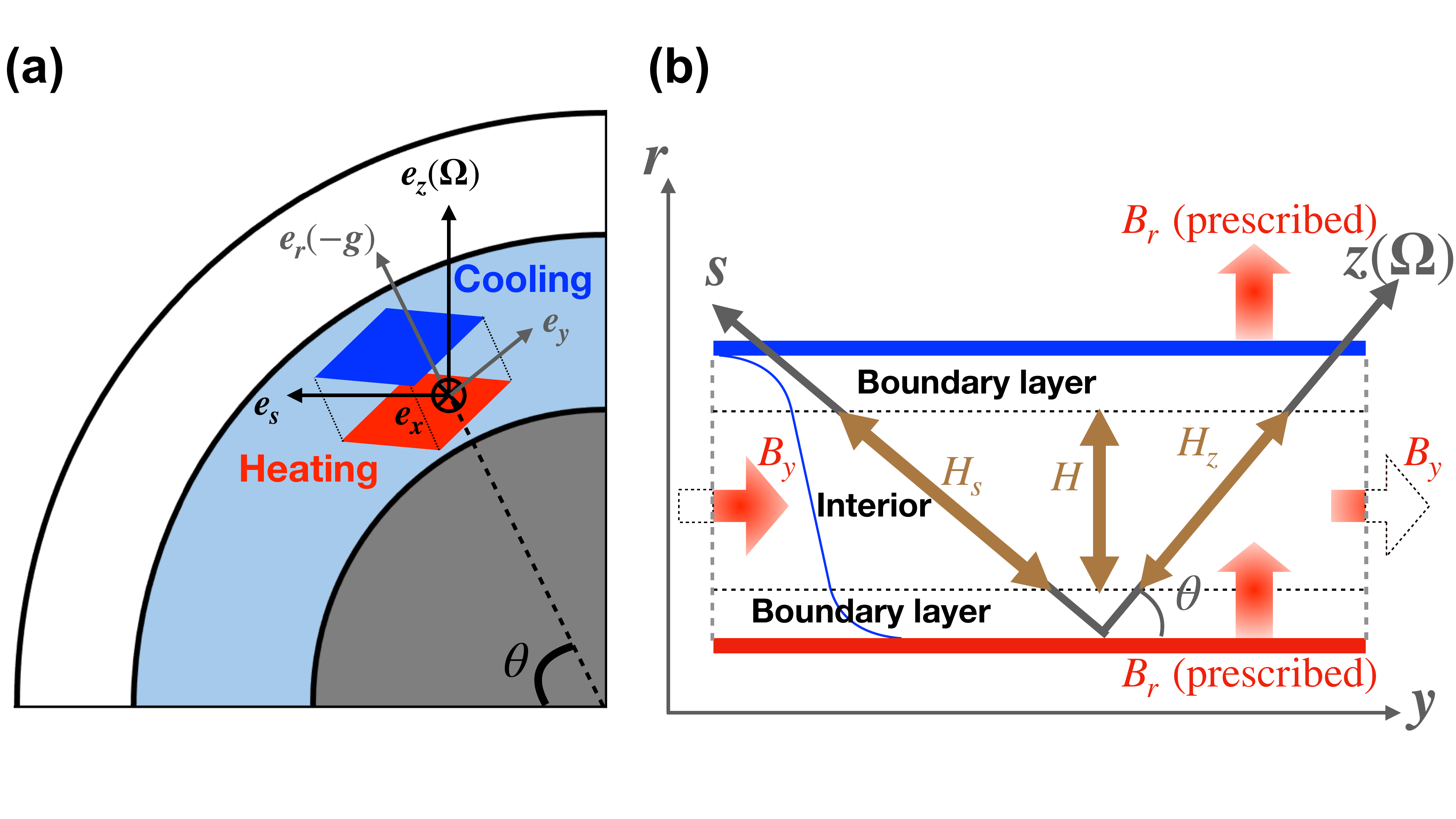}
    \caption{\textbf{Numerical simulation set up.} 
    (a) Schematic of the coordinate system. $\mathbf{e_x}$, $\mathbf{e_y}$, and $\mathbf{e_r}$ are the zonal, meridional, and vertical (opposite to gravity) directions, respectively. $\mathbf{e_z}$ is the axial direction aligned with planetary rotation, and $\mathbf{e_s}$ is the radial direction perpendicular to it. (b) Sketch of the numerical domain in the meridional-vertical plane. The vertical buoyancy flux $B_r$ is prescribed at the top and bottom boundaries. Horizontal boundaries are doubly periodic. The meridional buoyancy flux $B_y$ is not prescribed and is free to adjust. The domain is divided into an interior region and two boundary layers, distinguished by whether the temperature profile (sketched as the blue line) is linear or not (see Supplementary Text~S1 and Fig.~S1 for details). $H$, $H_z = H/\sin\theta$, and $H_s = H/\cos\theta$ represent the interior depth measured along the vertical, axial, and radial directions, respectively.}
    \label{fig:Model}
\end{figure}

We use the Massachusetts Institute of Technology General Circulation Model (MITgcm) \cite{adcroft2018mitgcm} to perform convection-resolving simulations in a local Cartesian box, with heating imposed at the bottom and cooling at the surface (Fig.~\ref{fig:Model}a). The simulations use the Boussinesq approximation, and the governing equations are:

\begin{equation}\label{eq:momemtum}
    \Dt{\bv} + 2 \bOmega \times \bv = -\nabla \Phi + b \eR + \nu \nabla^2 \bv,
\end{equation}

\begin{equation}\label{eq:buoyancy}
    \Dt{b} = \kappa \nabla^2 b,
\end{equation}

\begin{equation}\label{eq:continuity}
    \nabla \cdot \bv = 0,
\end{equation}

\noindent where $\bv$ is the velocity vector, $\nabla \Phi$ is the pressure gradient term, $b$ is the buoyancy, $\eR$ denotes the vertical direction (opposite to gravity), and $\nu$ and $\kappa$ are the eddy viscosity and diffusivity, respectively. We assume a linear equation of state with constant salinity: $b = g \alpha_T T$, where $T$ is the temperature anomaly from a constant reference, $\alpha_T=4\times 10^{-5}~\mathrm{K}^{-1}$, and $g=0.1$~m~s$^{-2}$.

Baroclinic eddies typically have much larger scales than slantwise convection, making it numerically infeasible to resolve both simultaneously. We therefore focus on the heat transport by slantwise convection in the absence of baroclinic instability. Equations~\ref{eq:momemtum}–\ref{eq:continuity} are solved in a local Cartesian domain with doubly periodic boundary conditions in the zonal ($\ex$) and meridional ($\ey$) directions, which removes any mean meridional temperature gradient and excludes the baroclinic instability associated with it. We apply free-slip boundary conditions, with a no-penetration bottom and a free surface top (a rigid lid at the top yields similar results; see Supplementary Text S1). A constant buoyancy flux $B_0$ is prescribed at both vertical boundaries (Fig.~\ref{fig:Model}b):

\begin{equation}\label{eq:boundary_condition}
    \kappa \pR{b} = B_0 \ \ \mathrm{at} \ \ r=0 \ \mathrm{and} \ L_r,
\end{equation}

\noindent where $r$ denotes the vertical direction and $L_r$ is the vertical domain extent.

We explore a parameter regime relevant to icy moon oceans (Table~\ref{tab:simulation}). These oceans lie in a transitional regime between rapidly rotating and non-rotating convection \cite{gastine2016scaling}, where rotation dominates the bulk interior dynamics but not the boundary layers and the smallest scales. To develop a parameterization for bulk heat transport by slantwise convection, we focus on the ocean interior, defined as the region where the horizontally and temporally averaged temperature gradient is approximately constant (Fig.~S1). We hypothesize that the CIA scaling law remains applicable in the ocean interior. To test this, we run simulations for four buoyancy fluxes ($B_0=10^{-14}$, $4\times10^{-14}$, $1.6\times10^{-13}$, and $6.4\times10^{-13}$~m$^2$~s$^{-3}$) and four rotation rates ($2\Omega=3.9\times10^{-5}$, $6.25\times10^{-5}$, $10^{-4}$, and $1.6\times10^{-4}$~s$^{-1}$), all relevant to the icy moon conditions (Table~S1). Due to the high computational cost of simulating the full ocean depth on icy moons (typically $10^4$–$10^5$~m), we test four reduced domain depths ($H_0=300$, 600, 1200, and 2400~m). We consider latitudes of $\theta = 15^\circ$, $30^\circ$, $60^\circ$, and $90^\circ$, excluding equatorial cases due to large temporal variability \cite{von2015generation}. The natural Rossby number \cite{jones1993convection,maxworthy1994unsteady,bire2022exploring}, $\mathrm{Ro^*} \equiv B^{1/2}(2\Omega)^{-3/2}H^{-1}$, in our simulations is within the estimated range for Titan, Europa and Ganymede, and about 10-100 times larger than the estimate for Enceladus (Tables 1 and S1). Importantly, the condition $\mathrm{Ro^*} \ll 1$ in both real icy moon oceans and our simulations confirms that they lie within the same regime, where convection is strongly influenced by rotation. All simulations are initialized with small-amplitude white noise in temperature and integrated until reaching a statistically steady state.

\begin{table}
\caption{Numerical simulation parameters.}
\centering
\begin{tabular}{c|cccccccc}
\hline
 Name  & $\theta$ ($^\circ$) & $B_0$ (m$^2$~s$^{-3}$) & $2\Omega$ (s$^{-1}$) & $H_0$ (m) & $N_x\times N_y \times N_r$ & $A_{\nu,\mathrm{smag}}$ & $\text{Ro}^*$ \\
\hline
  $L60_{\mathrm{Ctrl}}$  & $60$ & $4 \times 10^{-14}$ & $10^{-4}$ & 1200 & $140\times300\times300$ & $10^{-2}$ & $1.67 \times 10^{-4}$ \\
  $L60_{0.25B}$  & $60$ & $1 \times 10^{-14}$ & $10^{-4}$ & 1200 & $140\times300\times300$ & $10^{-2}$ & $8.33 \times 10^{-5}$ \\
  $L60_{4B}$  & $60$ & $1.6 \times 10^{-13}$ & $10^{-4}$ & 1200 & $140\times300\times300$ & $10^{-2}$ & $3.33 \times 10^{-4}$ \\
  $L60_{16B}$  & $60$ & $6.4 \times 10^{-13}$ & $10^{-4}$ & 1200 & $140\times300\times300$ & $10^{-2}$ & $6.67 \times 10^{-4}$ \\
  $L60_{0.39\Omega}$  & $60$ & $4 \times 10^{-14}$ & $3.9 \times 10^{-5}$ & 1200 & $140\times300\times300$ & $10^{-2}$ & $6.84 \times 10^{-4}$ \\
  $L60_{0.625\Omega}$  & $60$ & $4 \times 10^{-14}$ & $6.25 \times 10^{-5}$ & 1200 & $140\times300\times300$ & $10^{-2}$ & $3.37 \times 10^{-4}$ \\
  $L60_{1.6\Omega}$  & $60$ & $4 \times 10^{-14}$ & $1.6 \times 10^{-4}$ & 1200 & $140\times300\times300$ & $10^{-2}$ & $8.24 \times 10^{-5}$ \\
  $L60_{0.25H}$  & $60$ & $4 \times 10^{-14}$ & $10^{-4}$ & 300 & $140\times300\times300$ & $10^{-2}$ & $6.67 \times 10^{-4}$\\
  $L60_{0.5H}$  & $60$ & $4 \times 10^{-14}$ & $10^{-4}$ & 600 & $140\times300\times300$ & $10^{-2}$ & $3.33 \times 10^{-4}$ \\
  $L60_{2H}$  & $60$ & $4 \times 10^{-14}$ & $10^{-4}$ & 2400 & $140\times300\times300$ & $10^{-2}$ & $8.33 \times 10^{-5}$ \\
  $L15_{\mathrm{Ctrl}}$  & $15$ & $4 \times 10^{-14}$ & $10^{-4}$ & 1200 & $140\times300\times300$ & $10^{-2}$ & $1.67 \times 10^{-4}$\\
  $L15_{0.25B}$  & $15$ & $1 \times 10^{-14}$ & $10^{-4}$ & 1200 & $140\times300\times300$ & $10^{-2}$ & $8.33 \times 10^{-5}$\\
  $L15_{0.625\Omega}$  & $15$ & $4 \times 10^{-14}$ & $6.25 \times 10^{-5}$ & 1200 & $140\times300\times300$ & $10^{-2}$ & $3.37 \times 10^{-4}$\\
  $L15_{2H}$  & $15$ & $4 \times 10^{-14}$ & $10^{-4}$ & 2400 & $140\times300\times300$ & $10^{-2}$ & $8.33 \times 10^{-5}$\\
  $L30_{\mathrm{Ctrl}}$  & $30$ & $4 \times 10^{-14}$ & $10^{-4}$ & 1200 & $140\times300\times300$ & $10^{-2}$ & $1.67 \times 10^{-4}$\\
  $L30_{0.25B}$  & $30$ & $1 \times 10^{-14}$ & $10^{-4}$ & 1200 & $140\times300\times300$ & $10^{-2}$ & $8.33 \times 10^{-5}$\\
  $L30_{0.625\Omega}$  & $30$ & $4 \times 10^{-14}$ & $6.25 \times 10^{-5}$ & 1200 & $140\times300\times300$ & $10^{-2}$ & $3.37 \times 10^{-4}$\\
  $L30_{2H}$  & $30$ & $4 \times 10^{-14}$ & $10^{-4}$ & 2400 & $140\times300\times300$ & $10^{-2}$ & $8.33 \times 10^{-5}$\\
  $L90_{\mathrm{Ctrl}}$  & $90$ & $4 \times 10^{-14}$ & $10^{-4}$ & 1200 & $140\times300\times300$ & $10^{-2}$ & $1.67 \times 10^{-4}$\\
  $L90_{0.25B}$  & $90$ & $1 \times 10^{-14}$ & $10^{-4}$ & 1200 & $140\times300\times300$ & $10^{-2}$ & $8.33 \times 10^{-5}$\\
  $L90_{0.625\Omega}$  & $90$ & $4 \times 10^{-14}$ & $6.25 \times 10^{-5}$ & 1200 & $140\times300\times300$ & $10^{-2}$ & $3.37 \times 10^{-4}$\\
  $L90_{2H}$  & $90$ & $4 \times 10^{-14}$ & $10^{-4}$ & 2400 & $140\times300\times300$ & $10^{-2}$ & $8.33 \times 10^{-5}$\\
\hline
  $L60_{\mathrm{lowvisc}}$  & $60$ & $4 \times 10^{-14}$ & $10^{-4}$ & 1200 & $140\times300\times300$ & $10^{-3}$ & $1.67 \times 10^{-4}$\\
  $L60_{\mathrm{lowres}}$  & $60$ & $4 \times 10^{-14}$ & $10^{-4}$ & 1200 & $70\times150\times150$ & $10^{-2}$ & $1.67 \times 10^{-4}$\\
  $L60_{\mathrm{RigidLid}}$  & $60$ & $4 \times 10^{-14}$ & $10^{-4}$ & 1200 & $140\times150\times300$ & $10^{-2}$ & $1.67 \times 10^{-4}$\\
  $L60_{\mathrm{drag}}$  & $60$ & $4 \times 10^{-14}$ & $10^{-4}$ & 1200 & $140\times150\times300$ & $10^{-2}$ & $1.67 \times 10^{-4}$\\
  $L60_{0.5L_y}$ & $60$ & $4 \times 10^{-14}$ & $10^{-4}$ & 1200 & $140\times150\times300$ & $10^{-2}$ & $1.67 \times 10^{-4}$\\
  $L60_{2L_x0.5L_y}$  & $60$ & $4 \times 10^{-14}$ & $10^{-4}$ & 1200 & $280\times150\times300$ & $10^{-2}$ & $1.67 \times 10^{-4}$\\
\hline
\end{tabular}
\label{tab:simulation}
\end{table}

The typical plume size is estimated as $l_{\mathrm{plume}} = k B^{1/4} (2\Omega)^{-3/4} H^{1/2} \approx 100$~m using simulation parameters and an empirical pre-factor $k = 5$ \cite{fernando1989turbulent,jones1993convection,bire2022exploring}. To resolve these plumes, we use a grid spacing of 4~m in most simulations. We use 300 grid points in the meridional and vertical directions to ensure scale separation between the domain size and the plume scale, and use 140 points in the zonal direction to save computational costs. The shorter zonal domain relative to the meridional extent also promotes the formation of zonal rather than meridional jets \cite{julien2018impact,currie2020convection}, more realistically representing icy moon conditions \cite{soderlund2019ocean,ashkenazy2021dynamic,kvorka2022numerical,bire2022exploring,kang2023modulation,cabanes2024zonostrophic}. To test the effect of jets on heat transport, we also conduct sensitivity experiments with different horizontal aspect ratios ($L60_{0.5L_y}$ and $L60_{2L_x0.5L_y}$), as well as a simulation with linear bottom drag that suppresses jet formation ($L60_{\mathrm{drag}}$). For simulations with varying depth, we adjust the resolution to maintain a constant total number of grid points.

Fully resolving turbulence down to viscous scales is computationally infeasible. We therefore apply a Smagorinsky closure to parameterize subgrid-scale mixing, which assumes an isotropic forward energy cascade at the grid scale \cite{smagorinsky1963general,smagorinsky1993large}. Although the ``plume scale'' is resolved in our simulations, the characteristic scale at which rotation becomes important, $l_\Omega = B^{1/2}(2\Omega)^{-3/2}$, is approximately 0.2~m in our simulations, smaller than the grid spacing. This means that rotational effects may remain important below the grid scale, but are not adequately accounted for by the Smagorinsky closure. To test the sensitivity to the Smagorinsky closure, we performed simulations $L60_{\mathrm{lowvisc}}$ and $L60_{\mathrm{lowres}}$ to assess the effects of viscosity and resolution, and found no significant differences (see Supplementary Text S1). We set $\nu/\kappa=10$ to maintain minimally diffusive stable simulations. Based on the posterior-diagnosed Smagorinsky eddy diffusivity and viscosity, the simulation parameters fall within the transitional regime, consistent with that expected for icy moon oceans (Table~S2).

\section{Flux-gradient relation of the heat transport}\label{sec:scaling}

\begin{figure}[b!]
    \centering
    \includegraphics[width=1.0\linewidth]{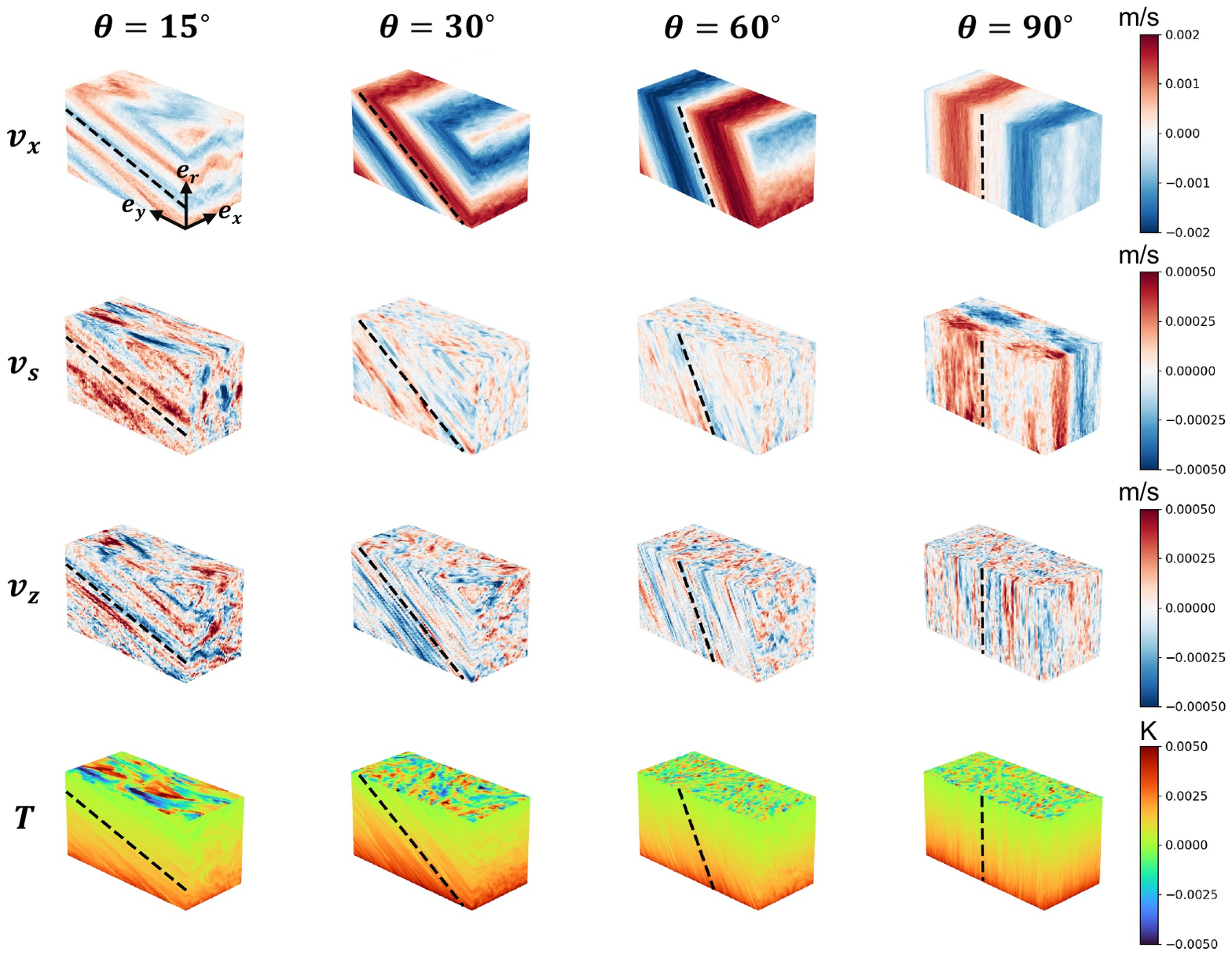}
    \caption{\textbf{Snapshots of the flow and temperature fields.} The first to fourth columns show results from the control simulations at different latitudes: $L15_{\mathrm{Ctrl}}$, $L30_{\mathrm{Ctrl}}$, $L60_{\mathrm{Ctrl}}$, and $L90_{\mathrm{Ctrl}}$, respectively. The first to fourth rows show the velocity components $v_x$, $v_s$, $v_z$ (c.f. Fig.~\ref{fig:Model}a), and the temperature anomaly $T$, respectively. In each panel, the front face shows a slice in the $y$–$r$ plane at $x=0$, the right face shows a slice in the $x$–$r$ plane at $y=L_y$, and the top face shows a slice in the $x$–$y$ plane at $r=L_r/2$, where $L_x$, $L_y$, and $L_r$ are the domain extents in the zonal, meridional, and vertical directions, respectively. Only the lower half of the domain is shown due to symmetry in the vertical direction, and to illustrate the flow and temperature fields in the interior. In the top face of the last row, the temperature anomaly is scaled up by a factor of 10 to enhance visibility. Dashed lines in the $y$–$r$ plane indicate the rotation axis ($\mathbf{e_z}$).}
    \label{fig:flow_fields}
\end{figure}

Strong zonal jets as well as slantwise convection plumes form in the simulations, all aligned with the planetary rotation axis (Fig.~\ref{fig:flow_fields}). Notably, the radial velocity component $v_s$ is comparable in magnitude to the axial component $v_z$ (Fig.~\ref{fig:flow_fields}, second and third rows), while both are roughly one order of magnitude smaller than the zonal jet velocities. This similarity ($v_s \sim v_z$) suggests that radial transport may be as important as axial transport. Moreover, the length scales of $v_s$ and $v_z$ are similar, indicating no clear scale separation between the two directions. This finding is supported by the heat transport spectra (Supplementary Text S2 and Fig.~S2), implying that both components should be treated concurrently in any parameterization of slantwise convection. Thermal boundary layers form near the top and bottom boundaries, maintaining a large temperature contrast between them (Fig.~S1). Temperature anomalies are also organized along the rotation axis, with horizontal temperature variations roughly an order of magnitude smaller than the full vertical temperature contrast (Fig.~\ref{fig:flow_fields}, last row).

\begin{figure}
    \centering
    \includegraphics[width=0.76\linewidth]{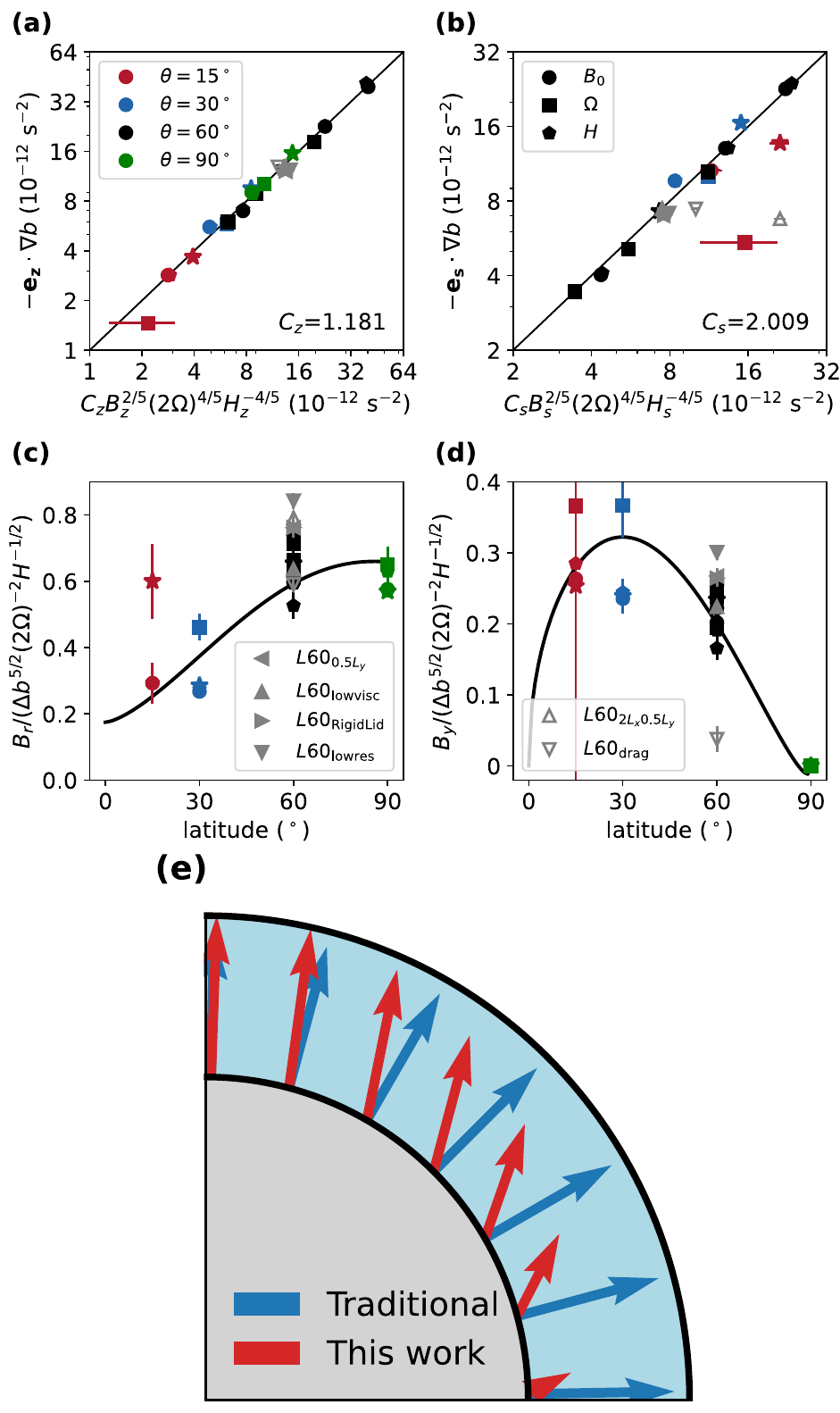}
    \caption{\textbf{Buoyancy flux–gradient relationships.} (a) and (b) show heat transport scaling in the axial and radial directions (Equations~\ref{eq:scaling-z} and \ref{eq:scaling-r}), respectively. (c) and (d) show the latitude dependence of normalized vertical ($B_r$) and meridional ($B_y$) buoyancy fluxes (Equations~\ref{eq:BR} and \ref{eq:By}), respectively. Marker color denotes latitude (legend in a), shape denotes parameter variations (legend in b), and grey markers denote sensitivity tests (legend in c–d). Filled/open markers represent simulations with/without zonal jets. Error bars show one standard deviation of temporal variability, typically negligible. Black lines denote the one-to-one line in (a) and (b) and the theoretical prediction in (c) and (d). Simulation $L60_{2L_x0.5L_y}$ is outside the range in (a) and (d); $L15_{0.625\Omega}$ is outside the range in (c). (e) illustrates the heat transport for a hypothetical icy moon ocean with uniform depth and vertical temperature contrast, with red and blue arrows showing the predicted heat flux from this study’s scaling laws versus a traditional convective adjustment scheme (see text for details). Arrow direction indicates the direction of heat flux, and arrow length is proportional to its magnitude.}
    \label{fig:scaling}
\end{figure}

We diagnose buoyancy gradients and fluxes from the simulations with the goal of informing parameterizations of bulk ocean heat transport in global models (Fig.~\ref{fig:scaling}). Specifically, we focus on the ocean interior, away from boundary layers, where the horizontally and temporally averaged temperature profile is approximately linear (Fig.~\ref{fig:Model}b). Within this region, we compute the buoyancy gradient $|\nabla b|$, the interior depth $H$, and the buoyancy flux $B$. The total buoyancy flux is decomposed into the axial component ($B_z$, parallel to the convective columns) and the radial component ($B_s$, perpendicular to the columns). The zonal heat flux is dynamically uninteresting because the averaged flux has to be approximately zero due to symmetry.

The CIA scaling reasonably captures the total heat transport magnitude (Fig.~S3). Motivated by this, we assume this scaling holds for heat transport along the convective plumes (i.e., the rotation axis) at fixed latitude. When latitude varies, the plumes tilt along the rotation axis, and their height changes to $H_z = H / \sin{\theta}$ (Fig.~\ref{fig:Model}b). We adjust $H$ to $H_z$ to account for the plume depth change, leading to the following flux–gradient scaling in the axial direction:

\begin{equation}\label{eq:scaling-z}
    -\mathbf{e_z} \cdot \nabla b \equiv \frac{\Delta b}{H_z} = C_z B_z^{2/5} (2\Omega)^{4/5} H_z^{-4/5}, \ H_z = \frac{H}{\sin{\theta}},
\end{equation}

\noindent where $C_z=1.181$ is an empirical constant obtained from simulations at $\theta = 60^\circ$ and applied to all latitudes. This scaling performs well for all simulations with zonal jets (filled markers), with errors below 10\% in most cases. The one notable exception is simulation $L15_{0.625\Omega}$ (red square), which exhibits large temporal variability in buoyancy flux and temperature profiles (Fig.~S4). Even in this case, the scaling predicts the temperature gradient within a factor of 2 (Fig.~\ref{fig:scaling}a). One sensitivity test ($L60_{2L_x0.5L_y}$, open upward triangle), which develops meridional rather than zonal jets, exhibits up-gradient axial heat transport and is not shown in Fig.~\ref{fig:scaling}a. However, because meridional jets are unlikely to occur in real icy moon oceans, we exclude this case from consideration here (see Section~\ref{sec:discussion} and Supplementary Text S3 for further discussion).

\begin{table}
\caption{Time-averaged simulation results.}
\centering
\setlength{\tabcolsep}{2pt}
\begin{tabular}{c|cccc}
\hline
 Name  & $\Delta b/H$~(s$^{-2}$) & $B_s$ (m$^2$~s$^{-3}$) & $B_z$ (m$^2$~s$^{-3}$) & $H$ (m) \\
\hline
  $L60_{\mathrm{Ctrl}}$ & $1.46 \times 10^{-11}$ & $7.64 \times10^{-15} \pm 3.94 \times 10^{-16}$ & $4.21 \times10^{-14} \pm 7.31 \times 10^{-16}$ & 872 \\
  $L60_{0.25B}$ & $8.06 \times 10^{-12}$ & $2.02 \times10^{-15} \pm 1.48 \times 10^{-16}$ & $1.04 \times10^{-14} \pm 2.88 \times 10^{-16}$ & 856 \\
  $L60_{4B}$ & $2.61 \times 10^{-11}$ & $3.35 \times10^{-14} \pm 5.23 \times 10^{-15}$ & $1.67 \times10^{-13} \pm 1.03 \times 10^{-14}$ & 880 \\
  $L60_{16B}$ & $4.54 \times 10^{-11}$ & $1.22 \times10^{-13} \pm 1.56 \times 10^{-14}$ & $6.72 \times10^{-13} \pm 2.99 \times 10^{-14}$ & 864 \\
  $L60_{0.39\Omega}$ & $6.89 \times 10^{-12}$ & $7.50 \times10^{-15} \pm 1.15 \times 10^{-15}$ & $4.15 \times10^{-14} \pm 2.26 \times 10^{-15}$ & 856 \\
  $L60_{0.625\Omega}$ & $1.02 \times 10^{-11}$ & $9.51 \times10^{-15} \pm 6.67 \times 10^{-16}$ & $4.08 \times10^{-14} \pm 1.34 \times 10^{-15}$ & 856 \\
  $L60_{1.6\Omega}$ & $2.10 \times 10^{-11}$ & $8.28 \times10^{-15} \pm 3.44 \times 10^{-16}$ & $4.19 \times10^{-14} \pm 8.12 \times 10^{-16}$ & 848 \\
  $L60_{0.25H}$ & $4.77 \times 10^{-11}$ & $9.09 \times10^{-15} \pm 1.41 \times 10^{-15}$ & $4.10 \times10^{-14} \pm 2.71 \times 10^{-15}$ & 220 \\
  $L60_{0.5H}$ & $2.62 \times 10^{-11}$ & $8.77 \times10^{-15} \pm 6.99 \times 10^{-16}$ & $4.07 \times10^{-14} \pm 1.47 \times 10^{-15}$ & 436 \\
  $L60_{2H}$ & $8.19 \times 10^{-12}$ & $8.35 \times10^{-15} \pm 2.62 \times 10^{-16}$ & $4.17 \times10^{-14} \pm 5.25 \times 10^{-16}$ & 1712 \\
  $L15_{\mathrm{Ctrl}}$ & $1.42 \times 10^{-11}$ & $3.37 \times10^{-14} \pm 6.94 \times 10^{-15}$ & $2.62 \times10^{-14} \pm 4.38 \times 10^{-15}$ & 928 \\
  $L15_{0.25B}$ & $1.10 \times 10^{-11}$ & $7.54 \times10^{-15} \pm 1.97 \times 10^{-15}$ & $1.16 \times10^{-14} \pm 1.41 \times 10^{-15}$ & 936 \\
  $L15_{0.625\Omega}$ & $5.36 \times 10^{-12}$ & $3.61 \times10^{-14} \pm 2.37 \times 10^{-14}$ & $1.45 \times10^{-14} \pm 8.70 \times 10^{-15}$ & 816 \\
  $L15_{2H}$ & $1.10 \times 10^{-11}$ & $2.97 \times10^{-14} \pm 6.85 \times 10^{-15}$ & $4.96 \times10^{-14} \pm 7.92 \times 10^{-15}$ & 1872 \\
  $L30_{\mathrm{Ctrl}}$ & $1.91 \times 10^{-11}$ & $1.75 \times10^{-14} \pm 2.65 \times 10^{-15}$ & $4.80 \times10^{-14} \pm 3.47 \times 10^{-15}$ & 928 \\
  $L30_{0.25B}$ & $1.11 \times 10^{-11}$ & $4.21 \times10^{-15} \pm 2.89 \times 10^{-16}$ & $1.25 \times10^{-14} \pm 3.92 \times 10^{-16}$ & 944 \\
  $L30_{0.625\Omega}$ & $1.61 \times 10^{-11}$ & $1.85 \times10^{-14} \pm 2.69 \times 10^{-15}$ & $4.69 \times10^{-14} \pm 3.97 \times 10^{-15}$ & 856 \\
  $L30_{2H}$ & $1.11 \times 10^{-11}$ & $1.67 \times10^{-14} \pm 1.05 \times 10^{-15}$ & $5.07 \times10^{-14} \pm 1.43 \times 10^{-15}$ & 1872 \\
  $L90_{\mathrm{Ctrl}}$ & $1.56 \times 10^{-11}$ & $-3.31 \times10^{-17} \pm 2.72 \times 10^{-18}$ & $4.09 \times10^{-14} \pm 1.11 \times 10^{-15}$ & 864 \\
  $L90_{0.25B}$ & $9.04 \times 10^{-12}$ & $-1.18 \times10^{-17} \pm 4.39 \times 10^{-19}$ & $1.04 \times10^{-14} \pm 1.60 \times 10^{-16}$ & 856 \\
  $L90_{0.625\Omega}$ & $1.02 \times 10^{-11}$ & $6.36 \times 10^{-18} \pm 4.18 \times 10^{-18}$ & $4.10 \times10^{-14} \pm 3.24 \times 10^{-15}$ & 864 \\
  $L90_{2H}$ & $8.86 \times 10^{-12}$ & $4.44 \times10^{-17} \pm 1.83 \times 10^{-18}$ & $4.16 \times10^{-14} \pm 7.96 \times 10^{-16}$ & 1680 \\
\hline
  $L60_{\mathrm{lowvisc}}$ & $1.49 \times 10^{-11}$ & $7.84 \times10^{-15} \pm 5.18 \times 10^{-16}$ & $4.19 \times10^{-14} \pm 9.90 \times 10^{-16}$ & 856 \\
  $L60_{\mathrm{lowres}}$ & $1.43 \times 10^{-11}$ & $7.64 \times10^{-15} \pm 4.82 \times 10^{-16}$ & $4.18 \times10^{-14} \pm 9.35 \times 10^{-16}$ & 784 \\
  $L60_{\mathrm{RigidLid}}$ & $1.40 \times 10^{-11}$ & $7.91 \times10^{-15} \pm 7.27 \times 10^{-16}$ & $4.12 \times10^{-14} \pm 1.37 \times 10^{-15}$ & 840 \\
  $L60_{\mathrm{drag}}$ & $1.49 \times 10^{-11}$ & $1.77 \times10^{-14} \pm 6.44 \times 10^{-16}$ & $3.57 \times10^{-14} \pm 1.17 \times 10^{-15}$ & 888 \\
  $L60_{0.5L_y}$ & $1.40 \times 10^{-11}$ & $7.85 \times10^{-15} \pm 7.22 \times 10^{-16}$ & $4.17 \times10^{-14} \pm 1.36 \times 10^{-15}$ & 848 \\
  $L60_{2L_x0.5L_y}$ & $1.36 \times 10^{-11}$ & $1.08 \times10^{-13} \pm 4.42 \times 10^{-16}$ & $-1.63 \times10^{-14} \pm 1.19 \times 10^{-14}$ & 864 \\
\hline
\end{tabular}
\label{tab:output}
\end{table}

To the best of our knowledge, no established theory exists for heat transport in the radial direction ($\mathbf{e_s}$, perpendicular to the convective plumes). We propose a semi-empirical scaling by analogy with the axial direction, projecting the depth along the radial direction as $H_s = H / \cos{\theta}$ (Fig.~\ref{fig:Model}b). This yields:

\begin{equation}\label{eq:scaling-r}
    -\mathbf{e_s} \cdot \nabla b \equiv \frac{\Delta b}{H_s} = C_s B_s^{2/5} (2\Omega)^{4/5} H_s^{-4/5}, \ H_s = \frac{H}{\cos{\theta}},
\end{equation}

\noindent where $C_s=2.009$ is fitted using simulations at $\theta = 60^\circ$. Alternative formulations using best-fit exponents in the natural Rossby number for heat transport in the radial direction ($\mathrm{Ro}^*_s=B_s^{1/2}(2\Omega)^{-3/2}H_s^{-1}$) were tested and found to yield similar performance (Fig.~S5). Notably, Equation~\ref{eq:scaling-r} implies that $H_s \to \infty$ at the pole, leading to $B_s \to 0$, which is consistent with the expected horizontal isotropy and agrees with simulation results showing $B_s/B_z \approx 0.1\%$ in all cases with $\theta = 90^\circ$ (Table~\ref{tab:output}). At lower latitudes, the tilt between the rotation axis and gravity breaks isotropy, generating correlated radial velocity and temperature anomalies that drive net radial heat transport (Fig.~S2). For most simulations, this scaling predicts radial heat flux with errors below 15\%.

Zonal jets influence the radial heat transport but have little effect on the heat transport along the rotation axis. This likely occurs because zonal jets are primarily barotropic along the rotation axis but have strong radial shear (Fig.~\ref{fig:flow_fields}). The radial shear deflects the flow, reducing the efficiency of radial heat transport \cite{aurnou2008convective,guervilly2017multiple}. Consequently, simulations without steady zonal jets exhibit stronger radial heat transport, producing a weaker buoyancy gradient than predicted. In simulations without zonal jets ($L60_{2L_x0.5L_y}$ and $L60_{\mathrm{drag}}$) and those with strongly meandering zonal jets ($L15_{\mathrm{Ctrl}}$ and $L15_{0.625\Omega}$; see Fig.~S6), the scaling tends to overestimate the buoyancy contrast (or equivalently, underestimate the buoyancy flux) in the radial direction. Even in these cases, the predicted values remain within a factor of 3 (Fig.~\ref{fig:scaling}b).

\section{Implications for the global ocean heat transport and ice shell thickness on icy moons}\label{sec:implication}

The flux–gradient scaling laws (Equations~\ref{eq:scaling-z} and \ref{eq:scaling-r}) predict the dependence of heat transport on latitude. The normalized vertical and meridional buoyancy fluxes are given by

\begin{equation}\label{eq:BR}
     \tilde{B}_r \equiv \frac{B_r}{\Delta b^{5/2} (2\Omega)^{-2} H^{-1/2}} = C_z^{-5/2} \sin^{3/2}{\theta} + C_s^{-5/2} \cos^{3/2}{\theta}
\end{equation}

\noindent and

\begin{equation}\label{eq:By}
    \tilde{B}_y = \frac{B_y}{\Delta b^{5/2} (2\Omega)^{-2} H^{-1/2}} = C_z^{-5/2} \sin^{1/2}{\theta} \cos{\theta} - C_s^{-5/2} \cos^{1/2}{\theta} \sin{\theta},
\end{equation}

\noindent respectively.

The vertical heat transport increases with latitude for fixed buoyancy contrast, rotation rate, and depth, as predicted by Equation~\ref{eq:BR} (Fig.~\ref{fig:scaling}c). This trend aligns well with our simulation results within a factor of 1.5, except for the lowest latitude cases ($L15_{\mathrm{Ctrl}}$ and $L15_{0.625\Omega}$). As discussed in Section~\ref{sec:scaling}, in these simulations, the jets are highly meandering, and the heat transport in the radial direction (close to vertical direction at low latitudes) is underestimated (Fig.~\ref{fig:scaling}b~\&~c).

The meridional heat transport is poleward and exhibits a non-monotonic dependence on latitude, peaking near $\theta = 30^\circ$ and vanishing at both the equator and the pole, as predicted by Equation~\ref{eq:By} (Fig.~\ref{fig:scaling}d). This pattern broadly agrees with the simulations within a factor of 1.5, although the scaling overestimates the meridional transport in simulation $L60_{\text{drag}}$, which does not develop jets. Only simulation $L60_{2L_x0.5L_y}$, which develops meridional jets, exhibits equatorward heat transport. This case is not shown in this panel and is excluded from consideration, as the meridional jets are not representative of icy moon ocean conditions (see Section~\ref{sec:discussion}).

For comparison with traditional convective parameterizations, which assume heat transport aligned with gravity, we evaluate the fluxes predicted by both the traditional scheme and the scaling laws developed in this study (Equations~\ref{eq:BR} and \ref{eq:By}), assuming constant ocean depth and vertical buoyancy contrast, with latitude taken at the bottom boundary. For the traditional scheme, we adopt the scaling from Equation~(3.10) of \citeA{gastine2016scaling}, which has previously been applied to icy moon ocean studies (e.g., \citeA{kang2023modulation}). Our scaling predicts weaker heat flux at lower latitudes and a heat flux vector that lies between the direction of gravity and the planetary rotation axis, resulting in a net poleward transport (Fig.~\ref{fig:scaling}e, red arrows). In contrast, traditional GCM convection schemes assume transport strictly aligned with gravity (Fig.~\ref{fig:scaling}e, blue arrows), thereby underestimating the poleward component. At the poles, our scaling converges with the traditional scheme, with a pre-factor difference of less than 10\%.

The predicted latitude dependence, with increased vertical flux at high latitudes and poleward meridional flux, suggests that slantwise convection may play an important role in redistributing heat toward the poles. If it dominates the heat transport, this redistribution would help maintain poleward-thinning ice shells on icy moons, qualitatively consistent with observations of Enceladus \cite{beuthe2016ice,vcadek2019long,hemingway2019enceladus}.

\section{Discussion}\label{sec:discussion}

We investigate heat transport associated with slantwise convection in the ocean interior, away from boundary layers, under conditions relevant to icy moons. Using convection-resolving simulations of rotating Rayleigh-Bénard convection on a tilted $f$-plane, we examine heat fluxes in both the axial and radial directions. We find scaling laws consistent with the classical diffusion-free (CIA) scaling at fixed latitude, extended here to include latitude dependence (Equations~\ref{eq:scaling-z}~\&~\ref{eq:scaling-r}). The heat transport in the axial and radial directions is of comparable magnitude, suggesting that both components should be included in parameterizations of slantwise convection in icy moon oceans. These scalings show that the total convective heat flux vector lies between the directions of gravity and planetary rotation. The vertical heat transport increases with latitude, and the meridional transport is poleward, both of which support the formation of a poleward-thinning ice shell, qualitatively consistent with observations of Enceladus.

Zonal jets emerge in most simulations. Sensitivity tests show that the presence of zonal jets influences the radial heat transport, although the jet strength itself has little effect. We performed one simulation with meridional jets, where the meridional heat flux reverses sign and becomes equatorward (Table~\ref{tab:output} and Supplementary Text S3). However, in realistic icy moon oceans, the $\beta$-effect arising from planetary rotation and spherical geometry promotes zonal jets and suppresses meridional jets \cite{rhines1975waves,vallis2017atmospheric}. We therefore focus on simulations with zonal jets only, where all simulations exhibit a poleward meridional heat transport.

Our simulations consider buoyancy driven solely by temperature. However, salinity can also contribute significantly to buoyancy stratification in icy moon oceans \cite{lobo2021pole,zeng2021ocean,kang2022does,zeng2024effect,ames2025ocean}. In the turbulent regime studied here, mixing is dominated by turbulence, which mixes heat and salt similarly. Therefore, as long as the equation of state is approximately linear for both temperature and salinity, our conclusions should remain valid in the presence of both buoyancy sources.

We adopt a Rayleigh–Bénard setup with uniform heating. However, seafloor heating on icy moons may be heterogeneous or spatially localized, as has been argued specifically for Enceladus \cite{choblet2017powering,bouffard2025seafloor}. Such localized heating could generate isolated plumes rather than global convection \cite{goodman2004hydrothermal,goodman2012numerical,kang2022ocean,bire2023divergent,zhang2025long,wang2025fate}. On Enceladus, these localized plumes are likely to lose buoyancy through mixing with the surrounding denser fluid, driven by baroclinic instability \cite{zhang2025long,wang2025fate}. Once buoyancy becomes horizontally homogenized, the system is expected to dynamically resemble the uniform heating case considered in this study. However, if highly localized heat sources exist on Europa, localized plumes there may or may not become well-mixed across the global ocean before reaching the surface \cite{wang2025fate}. A quantitative analysis of the behavior of isolated slantwise convecting plumes under a tilted rotation axis remains an important topic for future work.

Our work is designed to investigate processes in the ocean interior, and the boundary layer is not well resolved due to numerical constraints. This may influence the vertical velocity and buoyancy anomalies at the outer edge of the boundary layer and therefore introduces some uncertainty in the heat transport. We note, however, that boundaries in real icy moons are rough, and a viscous boundary layer along a smooth boundary is in any case unlikely to be applicable to icy moons. The possible role of realistic boundary layer processes remains an important area of research.

Our study isolates the effects of slantwise convection in the absence of baroclinic instability, which arises from large-scale meridional temperature gradients. Previous work has shown that baroclinic eddies tend to smooth out ice shell topography \cite{kang2022different,kang2022does,zeng2024effect}, in contrast to the poleward-thinning tendency driven by slantwise convection. \citeA{kang2023modulation} examined the interaction between these competing processes using a traditional, gravity-aligned convective parameterization. Revisiting this problem, using an improved representation of slantwise convection, remains an important direction for future investigation.

\appendix

\section*{Open Research Section}

The data on which this article is based are openly available in \citeA{zeng2025simulation}.

\section*{Conflict of Interest declaration}
The authors declare there are no conflicts of interest for this manuscript.

\acknowledgments
We thank Wanying Kang and Chang Liu for helpful discussions and comments. This work was completed using resources provided by the University of Chicago Research Computing Center. ChatGPT was used to help improve the clarity and grammar of the manuscript.

\bibliography{slantwiseconvection.bib}

\end{document}